\documentclass[10pt]{iopart}
\usepackage{iopams}  

\usepackage{graphicx}  
\usepackage{graphicx}  

\def\msun{\,{\rm M_\odot}}

\newcommand\be{\begin{equation}}
\newcommand\ee{\end{equation}}
\newcommand{\ba}{\begin{eqnarray}}
\newcommand{\ea}{\end{eqnarray}}

\begin{document}

\title[Gravitational Waves and Pulsar Timing]{Gravitational waves and pulsar timing: stochastic background, individual sources and parameter estimation}





\author{A Sesana$^1$ and \ A Vecchio$^2$\
}

\address{$^1$\ Center for Gravitational Wave Physics at the Pennsylvania State University, University park, State College, PA 16802, USA}

\address{$^2$\ School of Physics and Astronomy, The University of Birmingham, 
Edgbaston, Birmingham, B15 2TT, UK}

\begin{abstract}
Massive black holes are key ingredients of the assembly and evolution of cosmic structures. 
Pulsar Timing Arrays (PTAs) currently provide the only means to observe gravitational radiation from massive black hole binary systems with masses $\gtrsim 10^7\msun$. The whole cosmic population produces a signal consisting of two 
components: (i) a stochastic background resulting from the incoherent superposition of radiation from the all the sources, 
and (ii) a handful of individually resolvable signals that raise above the background level and are produced by sources sufficiently close and/or massive.
Considering a wide range of massive black hole binary assembly scenarios, we investigate both the level and shape of the background and the statistics of resolvable sources. We predict a characteristic background amplitude in the interval $h_c(f = 10^{-8}\,\mathrm{Hz}) \approx 5\times 10^{-16} - 5\times 10^{-15}$, 
within the detection range of the complete Parkes PTA. 
On average, at least one resolvable source produces  timing residuals that integrated over the typical time of observation lay in the range $\sim5-50$ ns.We also quantify the capability of PTAs of measuring the parameters of individual sources, focusing on the astrophysically more likely monochromatic signals produced by binaries in circular orbit. We investigate how the results depend on the number and distribution of pulsars in the array, by computing the variance-covariance matrix of the parameter measurements.
For 
plausible Square Kilometre Array (SKA) 
observations (100 pulsars uniformly distributed in the sky), and assuming a coherent signal-to-noise ratio of  10, the sky position of massive black hole binaries can be located within a 
$\approx 40$deg$^2$ error box, 
opening promising prospects for detecting a putative electromagnetic counterpart to the gravitational wave emission. The planned SKA, can plausibly observe these unique systems, although the number of detections is likely to be small. These observations would naturally complement on the high-mass end of the black hole distribution function future surveys carried out by the {\it Laser Interferometer Space Antenna} ({\it LISA}).
\end{abstract}

\section{Introduction}
Massive black hole (MBH) binary (MBHB) systems with masses in the range $\sim 10^4-10^{10}\msun$ are amongst the primary candidate sources of gravitational waves (GWs) at $\sim$ nHz - mHz frequencies ~\cite{enelt, jaffe, uaiti, ses04, ses05}. The frequency band $\sim 10^{-5}\,\mathrm{Hz} - 1 \,\mathrm{Hz}$ will be probed by the {\it Laser Interferometer Space Antenna} ({\it LISA}, ~\cite{bender98}), a space-borne GW laser interferometer being developed by ESA and NASA. The observational window $10^{-9}\,\mathrm{Hz} - 10^{-6} \,\mathrm{Hz}$ is already accessible with Pulsar Timing Arrays (PTAs;  e.g. the Parkes radio-telescope, ~\cite{man08}). PTAs exploit the effect of GWs on the propagation of radio signals from a pulsar to the Earth ~\cite{saz78, det79, ber83}, producing a characteristic signature in the time of arrival (TOA) of radio pulses.  The timing residuals of the fit of the actual TOAs of the pulses to the TOAs predicted by a given pulse emission and propagation model carry the physical information about unmodelled effects, including GWs  ~\cite{hel83, jen05}. The complete Parkes PTA  ~\cite{man08}, the European Pulsar Timing Array ~\cite{jan08}, and NanoGrav ~\cite{jen09} are expected to improve considerably on the capabilities of these surveys, and the planned Square Kilometer Array (SKA; {\it www.skatelescope.org}) will produce a major leap in sensitivity. 

Popular scenarios of MBH formation and evolution ~\cite{vhm03, kz06, mal07, yoo07} predict the existence of a large number of MBHBs emitting in the frequency range between $\sim 10^{-9}$ Hz and $10^{-6}$ Hz. PTAs can gain direct access to this population, addressing a number of unanswered questions in astrophysics (such as the assembly of galaxies and the physics of the relevant dynamical processes in galactic nuclei), by detecting gravitational radiation of two forms: (i) the stochastic GW background produced by the incoherent superposition of radiation from the whole cosmic population of MBHBs and (ii) GWs from individual sources that are sufficiently bright (and therefore massive and/or close) so that the gravitational signal stands above the root-mean-square (rms) value of the background. Both classes of signals are of great astrophysical and cosmological interest. In particular, the extraction of source parameters from the PTA observation may provide useful information about the system, helping in the identification of a putative electromagnetic counterpart.  

In this paper, we build on the results 
presented in ~\cite{papI, papII}, review the features of the signal emitted by the population of MBHB and investigate how accurately the parameters of individual binary systems can be measured. 
The plan of the paper is as follow. In Section 2 we describe the GW signal relevant to PTAs, treating separately the unresolved background and the resolvable sources. In Section 3 we consider 
several representative MBHB population models for generating the GW signals. Results, in terms of signal characterization and detectability, are discussed in Section 4, and in Section 5 we report preliminary results 
regarding the parameter estimation of individual sources. 
We briefly summarized our main findings in Section 6. 

\section{Gravitational wave signals for pulsar timing arrays}

A detailed derivation of the GW signal generated by a cosmological population of MBHBs is given in ~\cite{papI}. In the following we will assume each source to be a circular binary defined by a chirp mass ${\cal M}=M_1^{3/5}M_2^{3/5}/(M_1+M_2)^{1/5}$ ($M_2<M_1$ are the masses of the two black holes) and a restframe emitted frequency $f_r=(1+z)f$ ($f$ is the {\it observed} frequency and $z$ is the binary redshift), which is twice the orbital frequency. Consider now a MBHB population described by the distribution $d^3N/dzd{\cal M} d{\rm ln}f_r$, i.e. the comoving number of binaries emitting in a given logarithmic frequency interval with chirp mass and redshift in the range $[{\cal M},{\cal M}+d{\cal M}]$ and $[z, z+dz]$, respectively. The characteristic amplitude $h_c$ of the GW signal is given by: 
\begin{equation}
h_c^2(f) =\int_0^{\infty} 
dz\int_0^{\infty}d{\cal M}\, \frac{d^3N}{dzd{\cal M} d{\rm ln}f_r}\,
h^2(f_r).
\label{hch2}
\end{equation}
where
\begin{equation}
h(f_r)={8\pi^{2/3}\over 10^{1/2}}{{\cal M}^{5/3}\over d_L(z)}f_r^{2/3}\,,
\label{eqthorne}
\end{equation}
is the leading Newtonian quadrupole order contribution to the total GW emission, whose sky and polarisation averaged strain amplitude is given by ~\cite{tho87}, and $d_L(z)$ is the luminosity distance to the source.


\subsection{The unresolved background}
In observations with PTAs, radio-pulsars are monitored weekly for periods of years. The relevant frequency band is therefore between $1/T$ -- where $T$ is the total observation time, assumed, unless otherwise specified, to be 5 years throughout this paper -- and the Nyquist frequency $1/(2\Delta t)$ -- where $\Delta t$ is the time between two adjacent observations --, corresponding to $5\times 10^{-9}$ Hz - few$\times10^{-7}$ Hz. The frequency resolution bin is $1/T$. 
Given a resolution frequency bin, we can introduce the concept of {\it stochastic background}. Whether the superposition of many deterministic signals should be effectively considered a stochastic signal depends on the frequency range and the observation time. Here we will consider the usual (simplified) criterion that the stochastic level of the signal is the amplitude at which the total number of sources contributing to the frequency bin of width $1/T$ centered at $f$ is $\gg 1$.

The source distribution $d^3N/dzd{\cal M} d{\rm ln}f_r$ in equation (\ref{hch2}) can be evaluated numerically from 
models of MBHB formation and evolution (see Section 3), and is typically a steep function of ${\cal M}$ (see figure 5 in ~\cite{papI}), with very few massive binaries. As a consequence, at each frequency, the signal 
consists of the superposition of many contributions from relatively light binaries, overwhelmed by rare massive systems. A pragmatic way to evaluate the stochastic contribution to the signal is the following. We generate 1000 Monte--Carlo realization of the signal, computing numerically the mean differential distribution of sources with respect to redshift, chirp mass and frequency, i.e., $d^3N/dzd{\cal M}df$. If we consider an observed frequency $\tilde{f}$, and integrate the number of sources emitting in the frequency interval $[\tilde{f},\tilde{f} + 1/T]$ in a given mass range $[\tilde{\cal M}, +\infty[$ and over all redshifts, for any given $\tilde{f}$, we can always find a $\tilde{\cal M}$ such that
\begin{equation}
\int_0^{\infty}dz\int_{\tilde{\cal M}}^{\infty} d{\cal M} \int_{\tilde{f}}^{\tilde{f}+1/T} df\, \frac{d^3N}{dzd{\cal M} df}=1\,.
\label{stochastic}
\end{equation}
The integral above identifies the sources in the population that do not contribute to the stochastic background, i.e., those massive rare MBHBs which are likely to be individually resolved, because there is on average less then one of them per frequency bin. The result depends on the observational time, since the longer the observation, the narrower the frequency bin and, accordingly, the lower the level of signal contribution that can be considered stochastic. Broadly speaking the stochastic level is 
obtained by introducing in equation (\ref{hch2}) an upper cut--off in the ${\cal M}$-integral at each frequency, according to the condition given by equation (\ref{stochastic}). The spectrum of the observable induced residuals in the array is then given by 
\be
\delta t_\mathrm{bkg}(f)=h_c(f)/(2\pi f).
\label{noise}
\ee

\subsection{Residuals from individually resolvable sources}
The angle-averaged optimal signal-to-noise ratio (SNR) at which a signal from a MBHB radiating at (GW) frequency $\approx f$ can be detected using a {\em single} pulsar is
\be
\langle \rho^2 \rangle= \left[\frac{\delta t_\mathrm{gw}(f)}{\delta t_\mathrm{rms}(f)}\right]^2\,.
\label{e:snr}
\ee
In the previous expression $\delta t_\mathrm{rms}(f)$ is the root-mean-square value of the noise level $\delta t_n$ at frequency $f$, including the instrument noise and the background contribution given by equation \ref{noise}, $\langle .\rangle$ represents the average over the source position in the sky and orientation of the orbital plane, and $\delta t_\mathrm{gw}(f)$ is the characteristic amplitude of the timing residual over the observation time $T$ defined as:
\be
\delta t_\mathrm{gw}(f) = \sqrt{\frac{8}{15}}\frac{{\cal M}^{5/3}}{d_L}\,(\pi f)^{-1/3}\sqrt{fT}\,,
\label{e:deltatgw}
\ee
where the numerical pre-factor comes from the angle average of the amplitude of the signal and the $\sqrt{fT}$ term takes into account 
the build-up of coherent signal-to-noise ratio with the number of cycles.
Equation~(\ref{e:snr}) is appropriate to describe observations using a single pulsar; 
adding coherently the residuals from several pulsars yields an increase in SNR proportional to the square-root of the number of pulsars used in the observations. We will use the characteristic amplitude of the residuals $\delta t_\mathrm{gw}$ to quantify the strength of a GW signal in PTA observations; $\delta t_\mathrm{gw}$ can be used to compute in a straightforward way the SNR, as a function of the noise level and number of pulsars in the array (all of which are quantities that do not depend on the astrophysical model), and therefore assess the probability of detection of sources in the context of a given MBHB assembly scenario. The statistics of individually resolvable sources is studied by running 1000 Monte-Carlo realisations of the whole population of MBHBs and by selecting only those sources whose characteristic timing residuals, given by equation (\ref{e:deltatgw}), exceed the stochastic background level as defined by equations (\ref{hch2}) and (\ref{stochastic}) in the previous section. 

\section{Models of massive black hole binary populations}
The MBHB distribution  $d^3N/dzd{\cal M} d{\rm ln}f_r$ is the only ingredient that is needed 
to compute both the background level and the statistics of the individual sources. We generate distributions of coalescing MBHBs using merging galaxy catalogs derived  from the on-line Millennium Run database. The Millennium Simulation ~\cite{spr05} covers a volume of $(500/h_{100})^3$ Mpc$^3$ (here $H_{100}$ is the value of the Hubble parameter today normalised to 100 km$/$s$/$Mpc) and is the ideal tool to construct a statistically representative distribution of massive low/medium--redshift sources (which were demonstrated to dominate the signal in the PTA window \cite{papI}). As a first step we compile catalogues of galaxy mergers from the semi-analytical model of Bertone et al. ~\cite{ber07} applied to the Millennium run. We need to associate to each merging galaxy in our catalogue a central MBH, according to some sensible prescription. The Bertone et al. catalogue contains many properties of the merging galaxies, including the bulge mass $M_{\rm bulge}$, and the bulge rest frame magnitude $M_V$ both of the progenitors and of the merger remnant. This is all we need in order to assign a MBH to each galaxy. The process is twofold.

\begin{enumerate}
\item We populate the coalescing galaxies with
MBHs according to four different MBH-host prescriptions:
\begin{itemize}
\item $M_{\rm BH}-M_{\rm bulge}$ in the version given by Tundo et al. (~\cite{tun07}, ``Tu'' models);
\item $M_{\rm BH}-M_{\rm bulge}$, with a redshift dependence in the version given by Mclure et al. (~\cite{mcl06}, ``Mc'' models);
\item $M_{\rm BH}-M_V$ as given by Lauer et al. (~\cite{lau07}, ``La'' models);
\item $M_{\rm BH}-\sigma$ as given by Tremaine et al. (~\cite{tre02}, ``Tr'' models);
\end{itemize}
To each merging system we assign MBH masses according to one of the selected models so that we have the masses of the two MBH progenitors, $M_1$ and $M_2$. For each prescription we also calculate the mass of the MBH remnant, $M_r$, using the same relations. In all cases, the remnant mass is $M_r>M_1+M_2$, consistent with the fact that MBHs are expected to grow predominantly by accretion during the merger process.

\item For each MBH-host relation we consider three different accretion scenarios:
\begin{itemize}
\item The masses of the coalescing MBHs are $M_1$ and $M_2$; i.e., accretion is triggered {\it after} the MBHB coalescence. We label this accretion mode as "NA" (no accretion).
\item Accretion is triggered {\it before} the MBHB coalescence and only the more massive MBH ($M_1$) accretes mass; in this case the masses of the coalescing MBHs are $\alpha M_1$ and $M_2$, where $\alpha=(M_r-M_2)/M_1-1$. We label this accretion mode as "SA" (single BH accretion).
\item Accretion is triggered {\it before} the MBHB coalescence and both MBHs are allowed to accrete the same fractional amount of mass; in this case the masses of the coalescing MBHs are $\beta M_1$ and $\beta M_2$, where $\beta=M_r/(M_1+M_2)-1$. We label this accretion mode as "DA" (double BH accretion).
\end{itemize}
\end{enumerate}

Assigning a MBH to each galaxy, we obtain a list of coalescences (labelled by MBH masses and redshift). Each event  in the list is then properly weighted over the observable volume shell at each redshift to obtain the distribution $d^3N/d{\cal M}dzdt_r$, which is finally converted into $d^3N/dzd{\cal M} d{\rm ln}f_r$ using the standard quadrupole formula relation for the frequency shift $df_r/dt_r$ (equation (8) in \cite{papI}). 

\section{Results}
\begin{figure}
   \centering
   \includegraphics[width=3.5in]{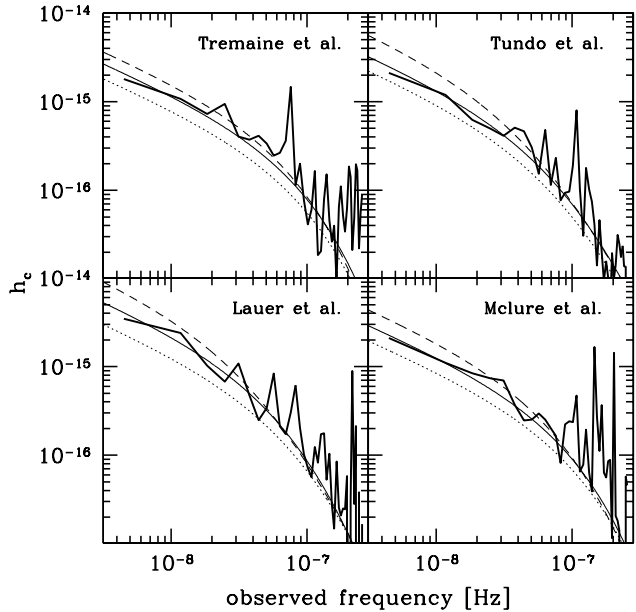}
      \caption{The characteristic amplitude of the GW signal from the population of MBHB systems. In each panel the thin lines identify the estimated level of the stochastic contribution assuming "SA" (solid line), "DA" (dashed line) and "NA" (dotted line) accretion modes. The total GW amplitude from a single Monte--Carlo realisation of the signal corresponding to the "SA" accretion mode is also shown as a thick solid line.
             }
         \label{fig1}
   \end{figure}

Single Monte--Carlo realizations of the typical signal, in terms of $h_c(f)$, are plotted in figure \ref{fig1} for the four selected MBHB population models, assuming the "SA" accretion mode. The signal spectrum can be decomposed in a ``confusion component" (the {\it stochastic} contribution), plus a 
handful of spikes due to particularly bright sources that are individually resolvable above the background level. In the following, we characterize both contributions. 

\subsection{Background: characterization and observability}
\begin{figure}
   \centering
   \includegraphics[width=3.5in]{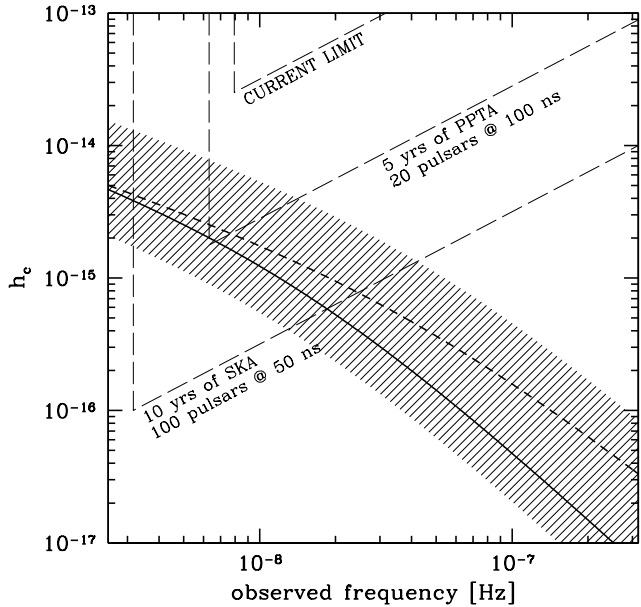}
      \caption{The sensitivity of PTAs to a GW stochastic background. The long--dashed thin lines represent the sensitivities of current and future timing experiments, as labeled in the figure. The thick lines depict the stochastic signal predicted by two representative MBH evolution models, see equation~(\ref{hfit}). The shaded area marks the possible range of GW background level, according to the model uncertainties.
             }
         \label{fig2}
   \end{figure}

The background level, computed following the prescription given in Section 2.1,
deviates significantly from a simple $f^{-2/3}$ power-law (see, e.g., \cite{phi01}) for $f\gtrsim 10^{-8}$ Hz, and it is well fitted by a function of the form
\be
h_c(f)=h_0\left(\frac{f}{f_0}\right)^{-2/3}\left(1+\frac{f}{f_0}\right)^{\gamma}\,.
\label{hfit}
\ee 
with ${h_0}=(1.46\pm0.67)\times10^{-15}$, ${f_0}=3.72^{+1.52}_{-2.30} \times 10^{-8}\,\mathrm{Hz}$, and $\gamma=-1.08^{+0.03}_{-0.04}$.
In general, the slope of the stochastic contribution to $h_c$ starts to deviate from $-2/3$ at around $10^{-8}$ Hz, and becomes as steep as $\approx -1.5$. Most of the signal predicted by semi-analytical approaches (e.g. \cite{uaiti}) is actually not present because of the {\it discrete nature of sources} (see \cite{papI} for a detailed discussion). The errorbars are computed taking into account for uncertainties related to: (i) the galaxy merger rate at low redshift which determines the abundance of MBHBs, (ii) the MBH-host galaxy relation assumed in populating galaxies with MBHs and (iii) the adopted accretion mode. The uncertainty range can be then compared to the sensitivities of ongoing and planned PTAs, as shown in Figure \ref{fig2}. The current sensitivity (as presented in \cite{jen06}) is a factor of $\sim4$ above the most optimistic estimate of the background. However, the complete Parkes PTA (PPTA) should provide a sensitivity improvement of about an order of magnitude in amplitude; several scenarios and/or parameter values within a given model could yield a detection at this level. The planned SKA will provide a major leap in sensitivity: a monitoring of 100 millisecond pulsars for 10 yrs at a precision level of $\delta t_\mathrm{rms} \sim 50$ ns will allow us to study the background spectral features in the frequency range $3\times10^{-9} \lesssim\, f\, \lesssim\,{\rm few}\times10^{-8}$, enabling the complete characterization of the GW signal. 
  
   \begin{figure}
   \centering
   \resizebox{\hsize}{!}{\includegraphics[clip=true]{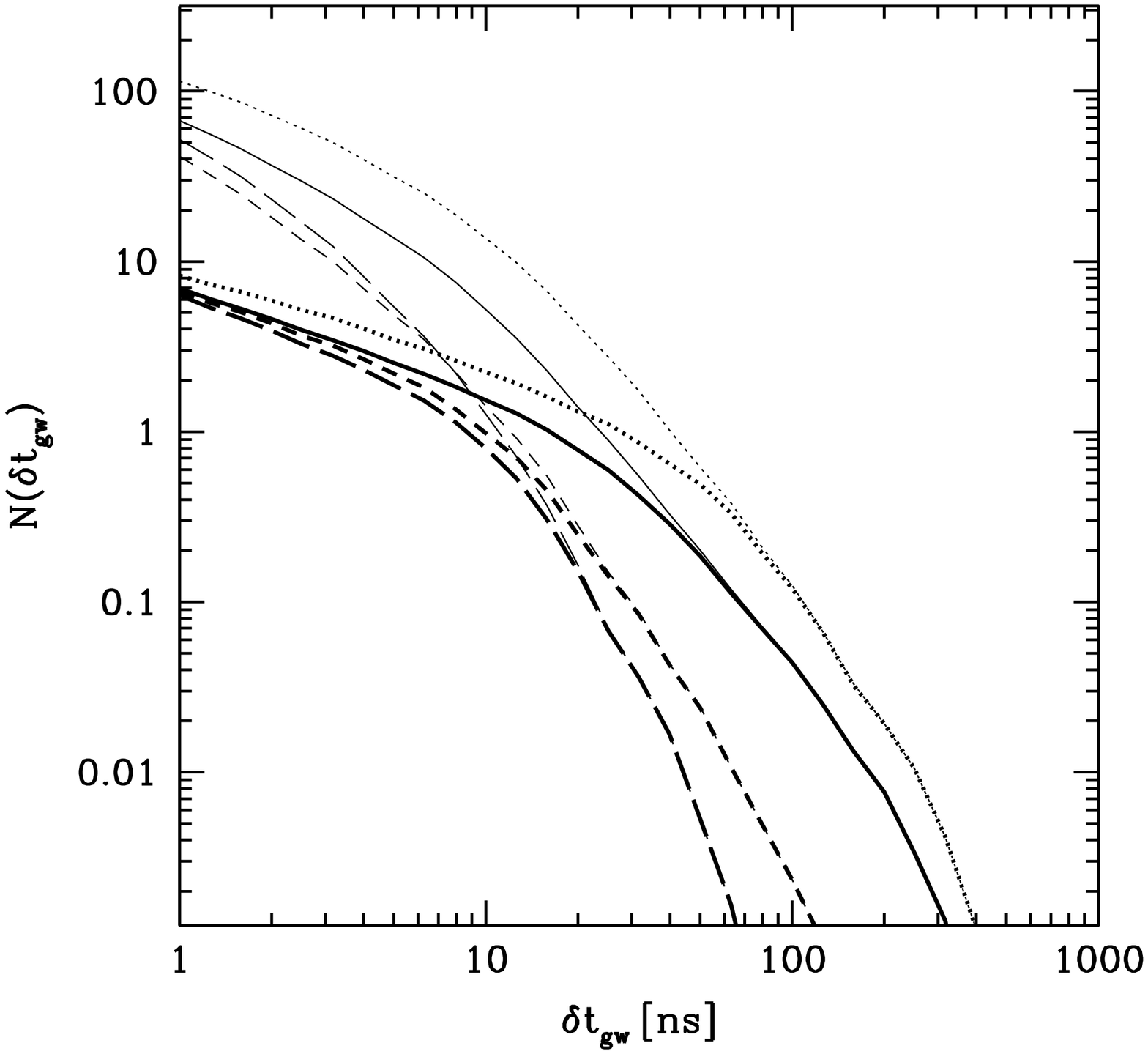}
   \includegraphics[clip=true]{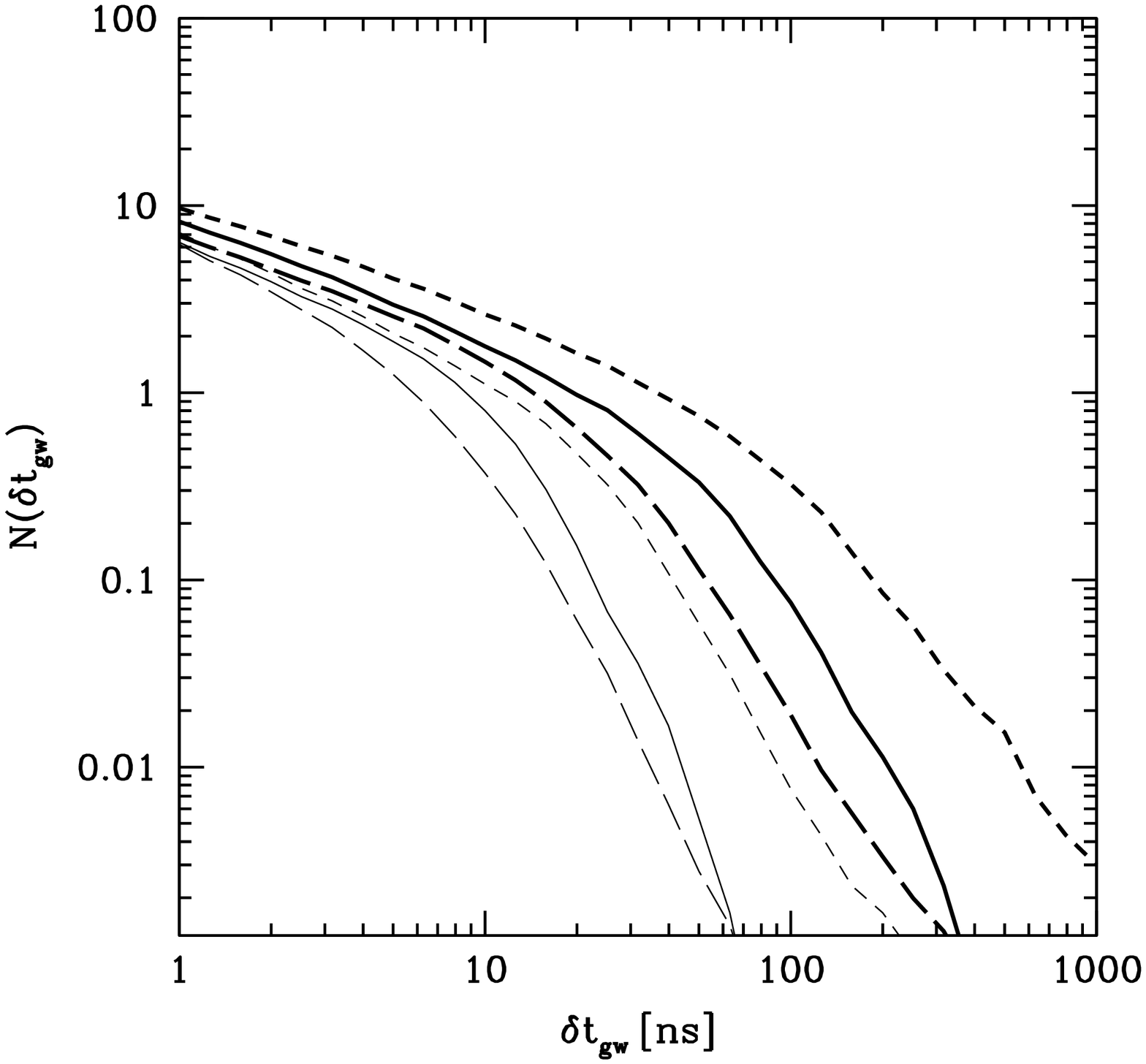}}
     \caption{{\it Left panel}: The effect of the MBH-host galaxy relation, assuming that accretion always takes place for a single black hole before merger ("SA" models), on the number of observable systems. The plot shows the number of total (thin lines) and resolvable (thick lines) sources $N(\delta t_\mathrm{gw})$ as a function of $\delta t_\mathrm{gw}$, see equation \ref{e:N}. Four different MBH merger scenarios are considered: Tu-SA (solid line), Tr-SA (long--dashed lines), Mc-SA (short--dashed lines) and La-SA (dotted lines), see Section 3 for a description of the models. {\it Right  panel}: The effect of the MBH accretion model on the number of individually observable systems. The plot shows the number of resolvable sources only, $N(\delta t_\mathrm{gw})$ as a function of $\delta t_\mathrm{gw}$. As reference for the MBH-host galaxy relation, models "La" (thick lines) and "Tr" (thin lines) are considered. The line style is as follow: model La-SA and Tr-SA (solid lines), La-DA and Tr-DA (short--dashed lines), and La-NA and Tr-NA (long--dashed lines). The duration of the observation is set to $T = 5$ yr .
}  
        \label{fig3}
    \end{figure}

\subsection{Properties and statistics of resolvable binaries}
To quantify the statistics of the individual sources, we cast the results in terms of the cumulative number of resolvable sources as a function of the timing residuals:
\be
N(\delta t_\mathrm{gw}) = \int_{\delta t_\mathrm{gw}}^\infty \frac{d N}{d(\delta t_\mathrm{gw}')}d(\delta t_\mathrm{gw}')\,,
\label{e:N}
\ee
where the integral is restricted to the sources that produce residuals above the rms level of the stochastic background.
Each Monte-Carlo realisation clearly yields a different value for $N(\delta t_\mathrm{gw})$
; the values shown in this section refer to the sample mean computed over the set of Monte-Carlo realisations.

In figure \ref{fig3}  we show the mean total number of individual sources that exceed a given level of timing residual (as defined by equation (\ref{e:N})), as a function of the timing residual. The left panel of figure \ref{fig3} shows results where the accretion prescription is the same ("SA"), but the underlying MBH-host relation changes;  assuming a sensitivity threshold of 30 ns, one expects to observe of the order of one source in the La-SA model, while there is only a probability $\approx$ 5\% for the Tr-SA model. The right panel shows that  different accretion scenarios may cause a fluctuation of a factor of $\approx 3$ in the mean number of expected  sources resolvable at a 10 ns level (e.g., between $0.5$ (Mc-NA) and $1.6$ (Mc-DA)). In turn, the timing precision required for positive detection is in the range $5-50$ns.

\section{Parameter estimation of resolvable systems}

\subsection{Fisher information matrix formalism}
If a source is individually resolved, then the observations provide a means to measure its astrophysical parameters. Most of the signals detected by PTAs may be from circular binaries exhibiting negligible frequency drift during the observation time. For this model, the timing residuals $r_\alpha(t;\vec{\lambda})$ from each pulsar (labelled by $\alpha$) depend on a 7-dimension parameter vector $\vec{\lambda}=\{R,\theta,\phi,\psi,\iota,f,\Phi_0\}$.
%
Here, $R\sim h/(2\pi f)$ is the amplitude, $\theta$ and $\phi$ describe the source position in the sky, $\psi$ is the source polarization angle, $\iota$ its inclination, $f$ its (constant) frequency and $\Phi_0$ its initial phase. Given an array of $N$ pulsar we can then compute the accuracy in the parameter estimation by computing the expected variance $\sigma_j^2 =  \left(\Gamma^{-1}\right)_{jj}$ of the parameter measurements, 
where $\Gamma_{jk}$ is the Fisher information matrix for the observation in the array of $\alpha=1,...,N$ pulsars, and is given by:    
\begin{equation}
\Gamma_{jk}=\sum_{\alpha=1}^N \Gamma_{jk}^{(\alpha)}\,=\sum_{\alpha=1}^N \left(\frac{\partial r_\alpha(t; \vec\lambda)}{\partial\lambda_j} \Biggl|\Biggr.\frac{\partial r_\alpha(t; \vec\lambda)}{\partial\lambda_k}
\right)\,.
\label{e:sigma}
\end{equation}
Here, $(x|y)$ denotes the noise weighted inner product of the two functions $x$ and $y$ -- in our analysis we assume that all the pulsars have the same noise level, which is modeled as Gaussian, stationary noise -- see \emph{e.g.}~\cite{cf94}. With this definition, $\Gamma_{\ln R,\ln R}=\rho^2$, where $\rho$ is the coherent PTA optimal signal-to-noise ratio. 
We normalized our calculation to a total SNR$=10$ and assume a ten years observation for this exercise. The accuracy in the parameter estimation scales linearly with the SNR.

   \begin{figure}
   \centering
   \resizebox{\hsize}{!}{\includegraphics[clip=true]{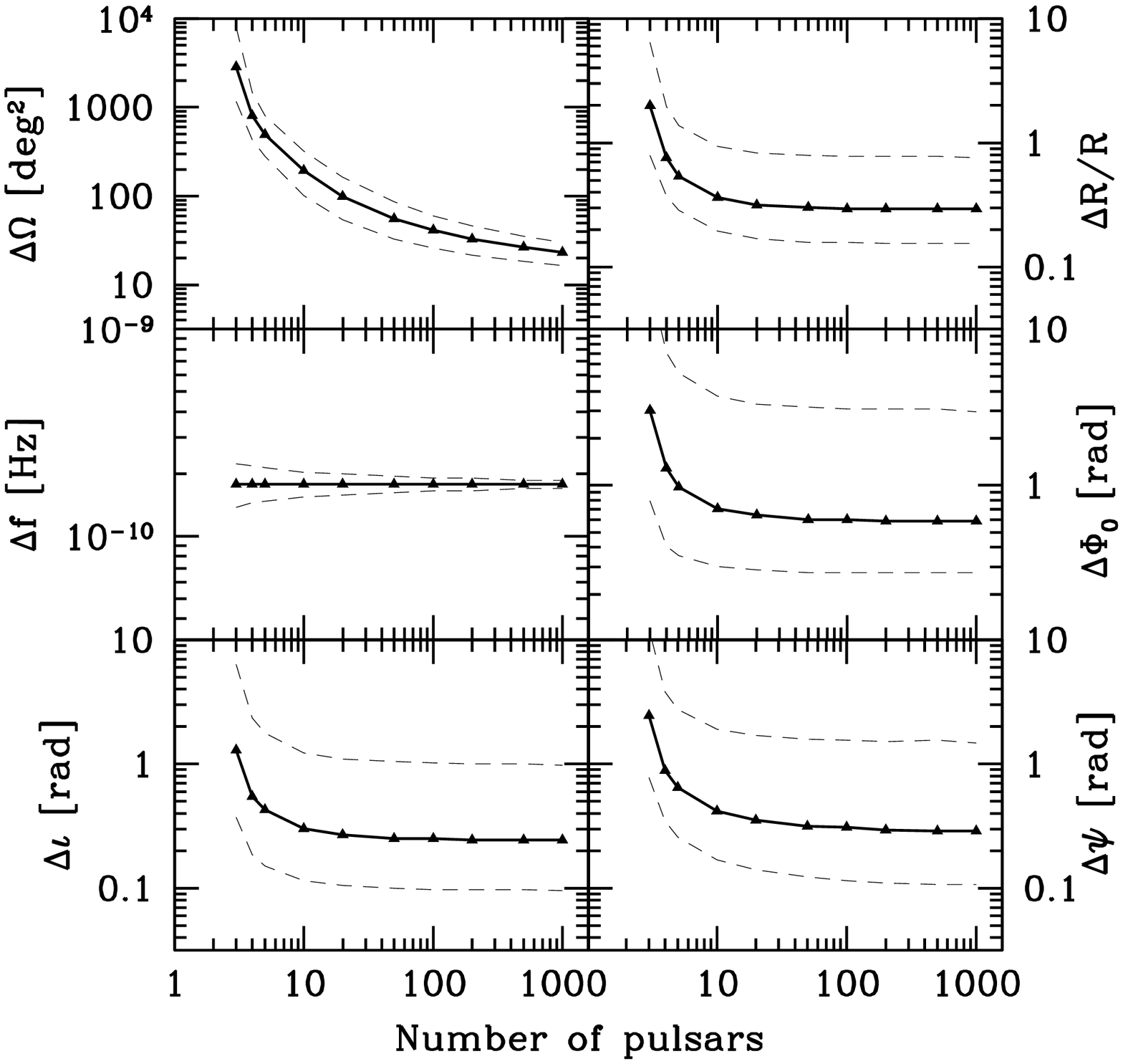}
   \includegraphics[clip=true]{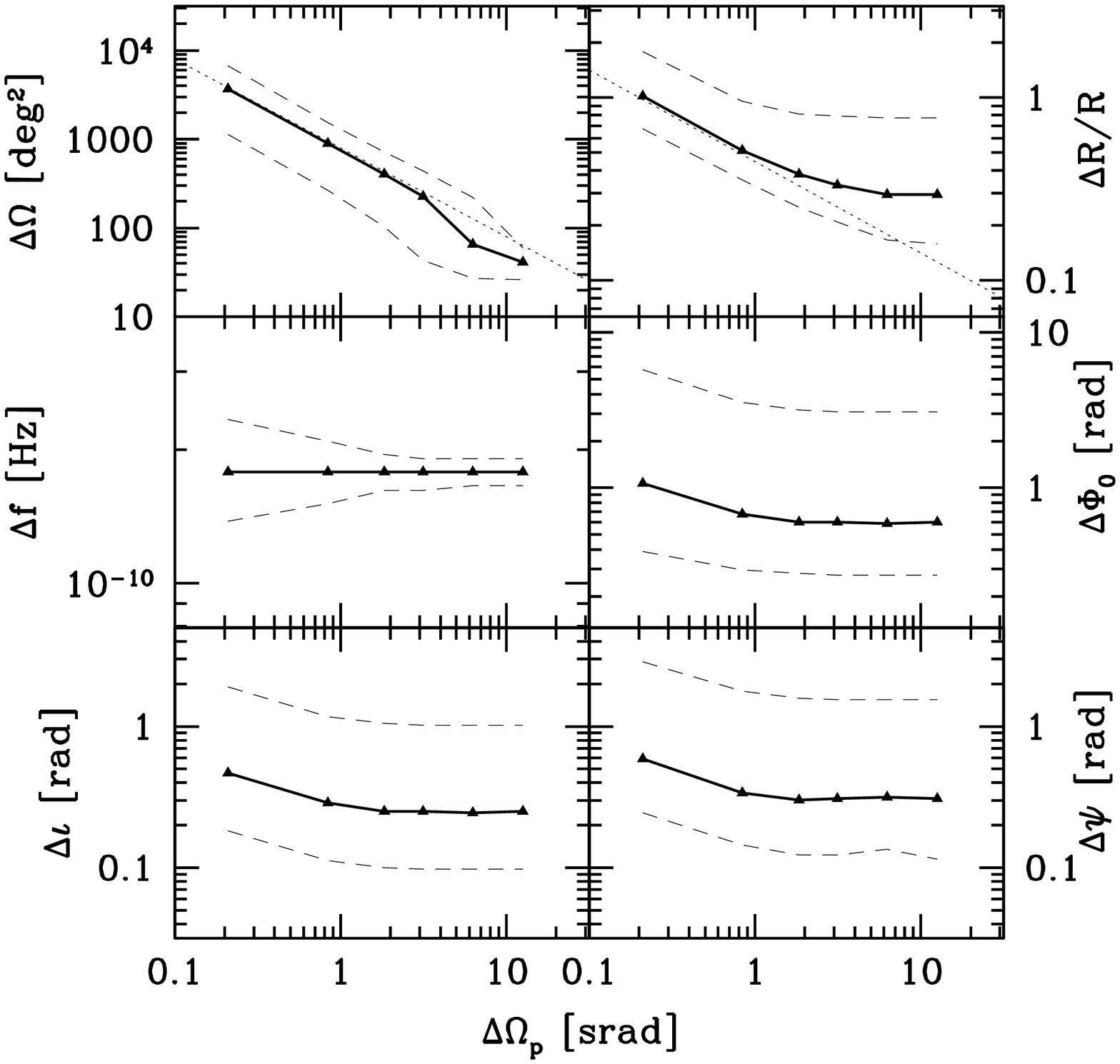}}
     \caption{{\it Left panel}: Median errors in the source parameter estimation. Each triangle is obtained averaging over $\ge 2.5\times 10^4$ Monte Carlo generated sources. In each panel, solid lines (triangles) represent the median errors, and the thin dashed lines label the 25$^{\rm th}$ and the 75$^{\rm th}$ percentile in the error distributions. {\it Left  panel}: errors as a function of the number of pulsars $M$, assuming a randomly distribution in the sky and a total SNR$=10$ in the array. {\it Right  panel}: Same as the left panel but now, in the $x$-axis we vary the sky coverage of the array in the range $[0,4\pi]$rad, keeping $N=100$ and SNR$=10$, linestyle as in the left panel.
}  
        \label{fig4}
    \end{figure}

\subsection{Results for an array of pulsar}
We consider two different observation scenarios: (i) we consider a PTA consisting of 
$N$ pulsars randomly and isotropically distributed in the sky varying $N=3, 5, 10, 20, 50, 100, 200, 500, 1000$; (ii) we fix $N=100$, and we change the sky coverage of the PTA 
(defined as the minimum solid angle containing all the pulsars) $\Delta\Omega_N=0.21, 0.84, 1.84, \pi, 2\pi, 4\pi$. For each pulsar distribution we generate $\ge 25000$ GW sources uniformly distributed in the sky with random initial phase and polarization, and inclination sampled according to a 
probability distribution $p(\iota)={\rm cos}(\iota)$ in the interval [$0,\pi$]. 
As a figure of merit of the performance of PTA we consider the 
median values of the parameter estimation accuracy as a function of $N$ and $\Delta\Omega_N$. Results are shown in figure \ref{fig4}, where we plot median and  25th and 75th percentile values of
$\sigma_{jj}$. 
At least three pulsars are necessary to constrain the source parameters. Assuming an isotropic distribution of pulsars (left panel), at a fixed SNR, the parameter determination accuracy improves a lot with increasing $N$ up to $\sim 20$. Adding further pulsars only improves the sky location of the source $\Delta\Omega=2\pi\sqrt{({\rm sin}\theta \Delta \theta \Delta \phi)^2-({\rm sin}\theta c^{\theta\phi})^2}$ ($c^{\theta\phi}$ is the $\theta-\phi$ correlation coefficient \cite{cf94}); its median value is well matched by the function $\Delta\Omega\approx 4\pi/(\rho^2\sqrt{N})$. Assuming a plausible SKA configuration of 100 pulsar isotropically distributed in the sky, a source with a signal-to-noise ration of 10 can be located within an error box $\approx 40\,\mathrm{deg}^2$; 
$\iota$ and $\psi$ are determined within a 0.3rad confidence. 
The frequency is pinned down to sub-frequency bin resolution, in this case $\approx 0.1\,\mathrm{nHz}$. 
The right panel highlights the beneficial effect of having pulsar isotropically distributed in the sky, and not concentrated in a limited area. 
Although the errors on most of the parameters are insensitive to the array sky coverage as long as $\Delta \Omega_N>1$ rad, the sky position accuracy increases linearly with the sky coverage over the whole range $[0,4\pi]$rad. Determining the source sky location is crucial to the task of finding a possible electromagnetic counterpart, which would have enormous scientific payoff. A uniform pulsar distribution in the array minimizes also the probability of detecting a binary in a region of the sky where the PTA sky resolution is poor.

\section{Summary}
We have explored the GW signal generated by a cosmological population of circular MBHBs, and its detectability with ongoing and planned PTAs. The signal spectrum can be decomposed in a stochastic contribution, 
on top of which rare, bright and individually resolvable sources sit. 
The stochastic background is well described by a double power law function. The slope of the $h_c$ spectrum starts to deviate from $-2/3$ at around $10^{-8}$ Hz, and becomes as steep as $\approx -1.5$. The current PTA sensitivity is a factor of $\approx 4$ above the most optimistic estimate of the background level. The complete Parkes PTA should guarantee a positive detection of the background for several selected scenario, while the planned SKA  will allow to study its spectrum in the frequency range $3\times10^{-9} \lesssim\, f\, \lesssim\,{\rm few}\times10^{-8}$. A timing precision of $5-50$ns (within the capabilities of the Parkes PTA and of the SKA) is required for detecting individual sources rising above the background. All the explored models predict at least 5-to-10 resolvable sources at a timing precision of 1ns. 
We have also evaluated the accuracy with which the parameters of a source can be measured, using an analysis based on the computation of the Fisher information matrix.
 For a plausible SKA configuration, with 100 pulsars uniformly distributed 
across the celestial sphere, the source sky location can be determined within an errorbox of $\approx 40\,\mathrm{deg}^2$, assuming a total signal-to-noise ratio of 10 in the array. The detection of these systems, along with a putative electromagnetic counterpart, would provide invaluable information for the study of MBHB pairing and evolution at low redshift and their role in the formation and evolution of massive galaxies and clusters.

\end{document}